\documentclass[final,5p,times,twocolumn]{elsarticle}

\usepackage{amssymb}
\usepackage{amsthm}
\usepackage{amsmath}
\usepackage{hyperref}
\usepackage{picture}
\usepackage{color}
\usepackage{nth}
\usepackage{gensymb}
\usepackage{lineno} 

\journal{Nuclear Instruments and Methods in Physics Research A }

\begin{document}

\begin{frontmatter}



\title{Conceptual design of electron beam diagnostics for high brightness plasma accelerator
}
 
\author[a]{A. Cianchi}\ead{alessandro.cianchi@uniroma2.it}
\author[b]{D. Alesini}
\author[b]{M. P. Anania}
\author[b]{F. Biagioni}
\author[b]{F. Bisesto}
\author[b]{E. Chiadroni}
\author[b]{A. Curcio}
\author[b]{M. Ferrario}
\author[b]{F. Filippi}
\author[b]{A. Ghigo}
\author[b]{A. Giribono}
\author[b]{V. Lollo}
\author[c]{A. Mostacci}
\author[b]{R. Pompili}
\author[b]{L. Sabbatini}
\author[b]{V. Shpakov}
\author[b]{A. Stella}
\author[b]{C. Vaccarezza}
\author[b]{A. Vannozzi}
\author[b]{F. Villa}

\address[a]{University of Rome Tor Vergata and INFN-Roma Tor Vergata, Via della Ricerca Scientifica 1, 00133 Rome, Italy}
\address[b]{INFN-LNF, Via Enrico Fermi 40, 00044 Frascati, Italy}
\address[c]{University La Sapienza of Roma, via Antonio Scarpa 24, 00133 Roma, Italy}


\begin{abstract}
A design study of the diagnostics of a high brightness linac, based on X-band structures, and a plasma accelerator stage, has been delivered in the framework of the EuPRAXIA$@$SPARC$\verb!_!$LAB project. In this paper, we present a conceptual design of the proposed diagnostics, using state of the art systems and new and under development devices. 
Single shot measurements are preferable for plasma accelerated beams, including emittance, while $\mu$m level and fs scale beam size and bunch length respectively are requested. 
The needed to separate the driver pulse (both laser or beam) from the witness accelerated bunch imposes additional constrains for the diagnostics. We plan to use betatron radiation for the emittance measurement just at the end of the plasma booster, while other single-shot methods must be proven before to be implemented. Longitudinal measurements, being in any case not trivial for the fs level bunch length, seem to have already a wider range of possibilities.  

\end{abstract}


\end{frontmatter}
\section{Introduction}
\label{sec:introduction}

Advancement in particle physics has always been linked with the availability of particle beams of ever increasing energy or intensity. 
Plasma-based concepts presently offer the highest gradient acceleration compared to other novel acceleration techniques like high-frequency W-band metallic RF structures, dielectric wakefield structures or direct laser acceleration. 
Plasma-based accelerators in fact replace the metallic walls of conventional RF structures with a plasma. 
This revolutionary change permits to avoid metallic or dielectric structure damage problems encountered in high-gradient operation. 
Laser beams (laser wakefield accelerator LWFA) or charged particle beams (particle wakefield accelerator, PWFA) may be adopted to excite space-charge oscillations in plasma. 
The resulting fields can be used for particle acceleration and focusing.
The EuPRAXIA facility \cite{eupraxia} (European Plasma Research Accelerator with eXcellence In Applications) is expected to be the first Research Infrastructure devoted to establish the scientific and technological basis required to build a compact and cost effective high energy (up to 5 GeV) machine based on plasma accelerator technology.
In order to be eligible for hosting this infrastructure, the INFN is considering the construction of a new LNF infrastructure, called EuPRAXIA$@$SPARC$\verb!_!$LAB \cite{eusparc}.
This facility by itself will equip LNF with a unique combination of a high brightness GeV-range electron beam and a 0.5 PW-class laser system and will allow to establish a FEL user community interested to exploit the proposed radiation source and the possible future extensions of the radiation spectrum, from the water window (4-2 nm) down to the Angstrom scale.

\begin{figure}[htb]
  \centering
  \includegraphics*[width=85mm]{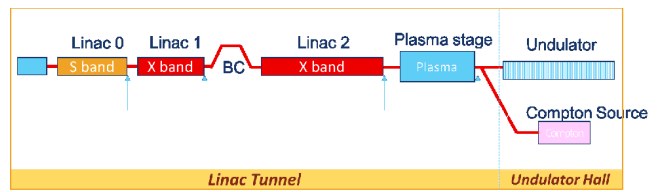}
  \caption{Machine schematic layout. A S-band photoinjector is followed by a X-band linac and a plasma booster section.}
  \label{fig1}
\end{figure}

The machine will be equipped by X-band RF accelerator structures plus a plasma booster. 
A sketch of the machine is shown in Fig.\ref{fig1}.
The final energy is foreseen to be in order of 1 GeV. 
It will produce electron beams with high brightness, short temporal duration and small transverse emittance. 
Moreover also plasma accelerated beams will be available in different schemes.

To cover all possible scenarios the beam diagnostics must be particularly advanced. 
For electron beam accelerated by the main linac, we can allow multi-shot measurements, mainly due to the inherent machine stability.

In table \ref{table1} are reported the beam transverse and longitudinal parameters for the bunch injected inside the plasma.

\begin{table}[!htbp]
  \centering
  \begin{tabular}{|r|c|c|} 
  \hline 
  & \textbf{Units} & \textbf{@plasma entrance} \\ 
  \hline 
  \textbf{Bunch charge} & pC & 30  \\ 
  \hline 
  \textbf{Bunch length rms}& fs & 12  \\ 
  \hline 
  \textbf{Peak current}& kA & 3  \\ 
  \hline 
  \textbf{Rep. rate}& Hz & 10  \\ 
  \hline 
  \textbf{Rms norm. emittance}& $\mu$m & 0.44 \\ 
  \hline
  \textbf{Transverse beam size (rms)}& $\mu$m & $\approx$ 1 \\ 
  \hline  
\end{tabular}
\caption{Beam parameters at plasma entrance}
 \label{table1}

\end{table} 

At the entrance of plasma booster the beam transverse dimension is about 1 $\mu$m rms, and its length is in the order of few fs. 
After the plasma acceleration, bunch length and emittance are quite preserved, but the inherent instability of the plasma acceleration demands also single shot measurements.

We will divide our analysis in three different branches: transverse, longitudinal, charge and trajectory diagnostics.
\section{Diagnostics Devices}

\subsection{Transverse diagnostics}
\begin{figure}[htb]
  \centering
  \includegraphics*[width=90mm]{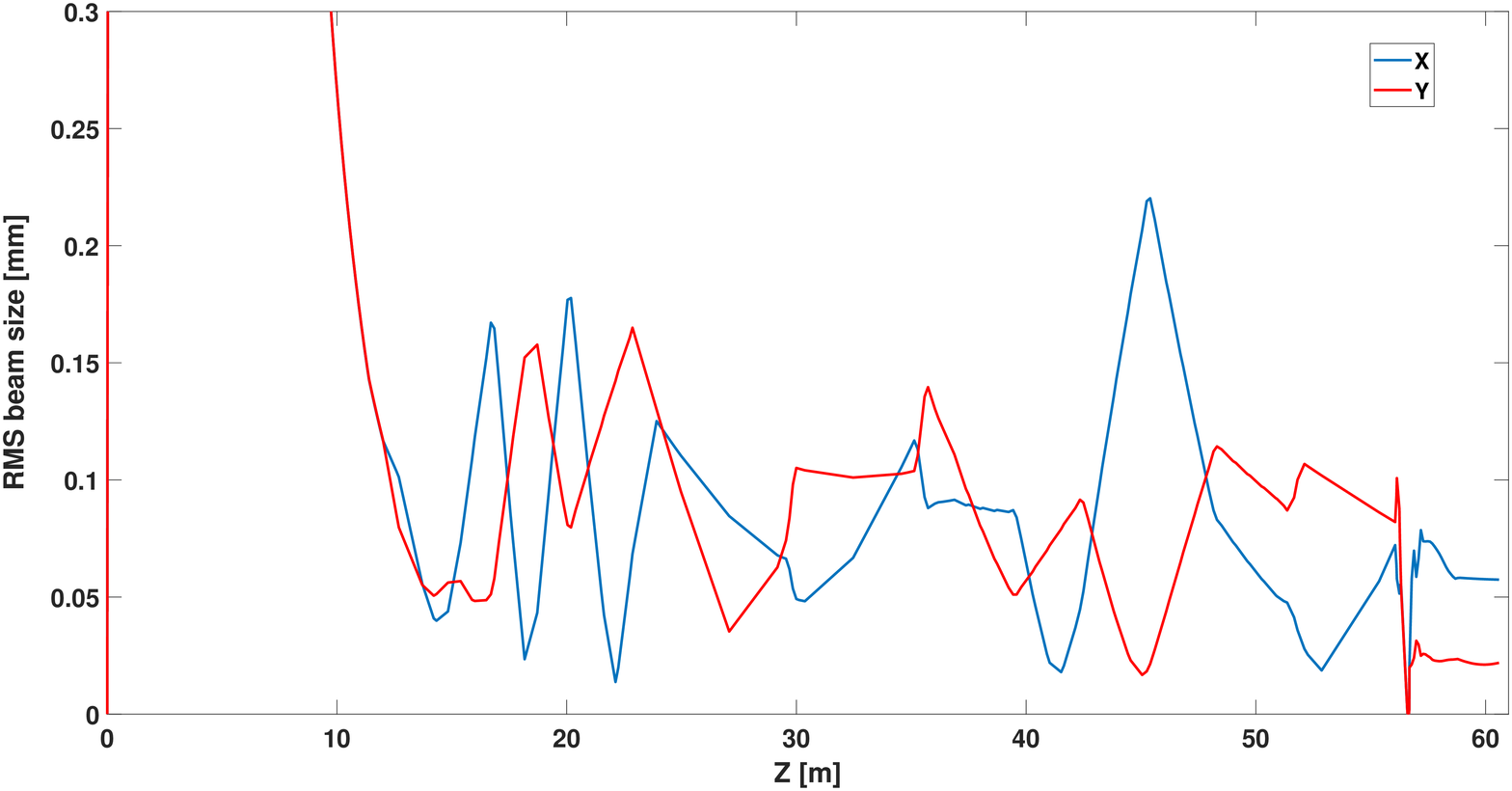}
  \caption{Beam envelop evolution along the linac. The transverse dimensions approach $\mu$m level at the plasma booster.}
  \label{fig2}
\end{figure}

There are two main measurements for transverse diagnostics: emittance and envelope. 
The envelope is very important in order to properly match the beam along the machine, comparing the measured dimensions with the simulated ones. 
In Fig.\ref{fig2} are shown the horizontal and vertical rms beam sizes. 

\begin{figure}[htb]
  \centering
  \includegraphics*[width=80mm]{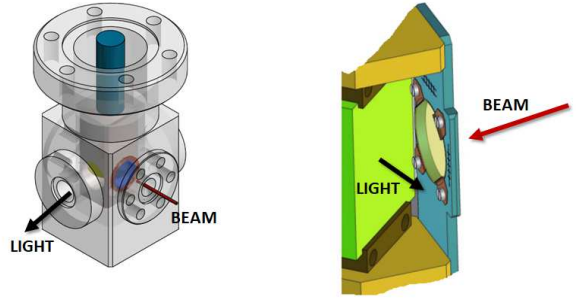}
  \caption{Example of a compact design for beam size measurements. The overall length is only 40 mm.}
  \label{fig3}
\end{figure}

Usually scintillator screens, like YAG:Ce, or Optical Transition Radiation (OTR) monitors are in use for such a task. 
In particular, YAG screens are widely used when the beam charge is below about few tens of pC, due to their better photon yield. 
In order to alleviate the problem of the depth of field and the crystal view angle \cite{Gero}, the conventional mounting considers to put the YAG normal to the beam line and a mirror placed at 45 degrees with respect to this direction to reflect the radiation at 90 degrees with respect to the beam line. 
This radiation is then extracted via a vacuum windows and imaged on a CCD. 
Compact design is required in order to preserve the compactness of the whole machine. 
As an example in Fig.\ref{fig3} is shown a new compact design that uses only 40 mm of longitudinal space to host this device.

The emittance measurement can be performed for the main linac with very well-known techniques, like quadrupole scan and multiple screens \cite{lohl2006measurements}. 
However for plasma accelerated beams the inherent shot to shot instabilities, with high beam divergence and the needs to separate driver and witness, prevent the use of such a diagnostics just after the plasma channel. 

For these reasons we consider the use of a different approach in order to measure the beam parameters only inside the plasma channel. 
No measurements will be performed just after the plasma, leaving space for capture optics. 
After that, innovative single shot measurement must be implemented. 

The energy spread must be also kept as low as possible because, following \cite{migliorati2013intrinsic}, even the 6D rms emittance is not preserved in a drift with energy spread, and so the measurement of the emittance is strongly dependent on the measurement position. 

The measurement inside the plasma bubble can be performed by means of betatron radiation. 
The diagnostics based on betatron radiation \cite{rousse2004production} has been developed in recent years in several laboratories, relying on the measurement of the spectrum, (for instance among the other see \cite{plateau2012low}) or on the diffraction from a knife edge \cite{kneip}. 

However, these systems were able to measure just the beam profile and divergence, neglecting the correlation term.
Only recently a new algorithm has been developed in order to retrieve the correlation term \cite{curcio2017trace}. 
By using the simultaneous measurement of the electron and radiation energy spectrum, it is possible to have a better reconstruction of the phase space. 
This method relies with some approximation on the initial phase space of the particle. But if the beam is external injected inside the plasma, the knowledge of the initial 6D phase space solves also this ambiguity.

\begin{figure}[htb]
  \centering
  \includegraphics*[width=80mm]{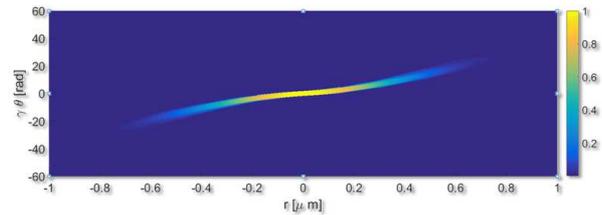}
  \caption{Reconstructed phase space with betatron radiation. Parameters are reported in \cite{curcio2017trace}.}
  \label{fig4}
\end{figure}

In Fig.\ref{fig4} is reported a reconstructed phase space with this technique.
In order to collect the betatron radiation, sooner or later we have to separate the radiation from the electron beam, with a dipole. 
Unfortunately the bending of the beam produces synchrotron radiation, and its spectrum can overlap with the betatron radiation.

\begin{figure}[htb]
  \centering
  \includegraphics*[width=80mm]{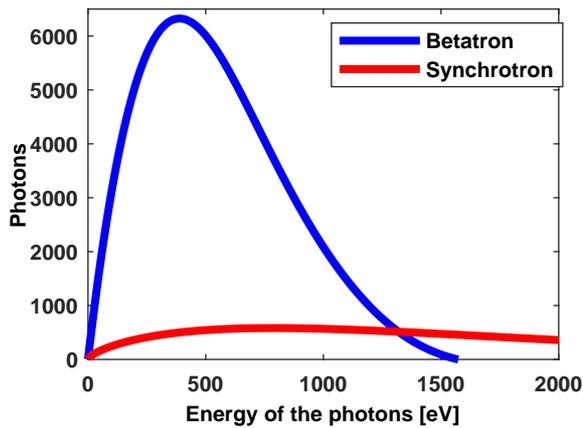}
  \caption{Comparison between Synchrotron radiation emitted by a bending magnet and Betatron radiation. Beam charge 30 pC, energy 1 GeV, plasma density 2 10$^{16}$ cm$^{-3}$; magnet filed 1.5 T, radius of curvature 2.2 m}
  \label{fig5}
\end{figure}

To this end we can notice that usually the betatron radiation is much stronger with respect to synchrotron radiation, and there is the additional degree of freedom in the bending angle. 
In Fig.\ref{fig5} is reported a comparison between the spectrum of the betatron radiation of 30 pC beam accelerated up to 1 GeV inside a plasma and the synchrotron radiation produced by a bending magnets. 
Details of the simulation are in the caption.

However, an open problem is related to the separation between the betatron radiation coming from the witness and from the driver in the beam driven scheme. 
In this case the driver contains much more charge with respect to the witness and so only a clear energy separation of the two spectra can solve the problem.
Obviously in the case of external injection this problem disappears.

After the capture optics another single shot emittance measurement is needed, for instance, in order to match the beam to the undulator. In our opinion there is not yet a clear solution to this problem. 
Pepper pot like techniques \cite{lejeune1980} are not easy to be implemented at high energy, because to clearly separate the beamlets a very tick target is needed. 
But the increase in the thickness reduces the angular acceptance of the beam, resulting in a likely cut of the phase space. 
However, the pepper pot suffers also for a problem related to the sampling nature of this measurement, as already discussed in \cite{cianchi2013}.

A more careful analysis of this problem reveals that part of the reduction in the sensitivity of this technique comes from the use of a double sampling, one on the target that selects beamlets, and one on the screen to image them. 
We have proposed and we are testing a new device \cite{cianchi2017transverse}, a sort of optical pepper pot, where there is only one sampling at the source level. 
It makes use of Optical Transition Radiation (OTR) produced when a charge passes through a metallic foil. 

	\begin{figure}[htb]
  \centering
  \includegraphics*[width=80mm]{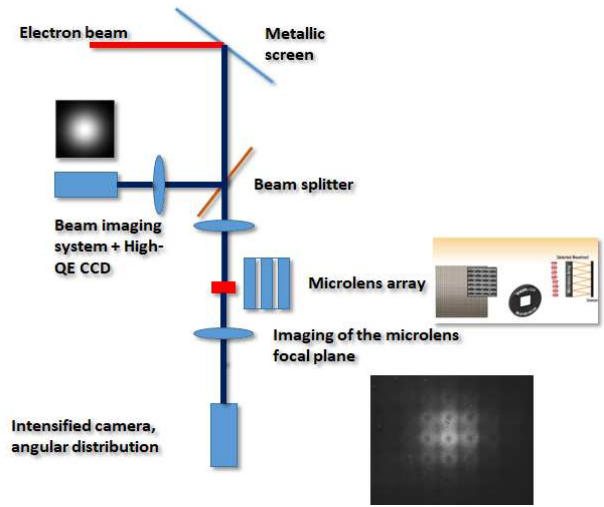}
  \caption{The OTR produced by the beam is divided in two arms. In one a camera record the beam spot size. In the other arm a replica of the radiation field on the source is produce in the front plane of a microlens array. The angular distribution is then retrieved for each microlen imaging their focal plane in a intensified camera.}
  \label{fig6}
\end{figure}

The angular distribution of the emerging radiation contains information about the angular divergence of the beam. 
By using an optical system to reproduce outside the vacuum chamber the source radiation field and sampling by means of a microlens array, it is possible to retrieve the value of the beam divergence in different transverse positions. 
Acquiring at the same time also the beam spot allows, in principle, the measurement of the emittance in single shot. 
In Fig.\ref{fig6} there is a sketch of the system, with also a picture of the measurement to show the output of the method. 

While we have already performed a test of this device we were not able so far to retrieve the value of the emittance. 
In fact the resolution increases quickly with the energy and it is too poor at actual SPARC energy value.
We consider only incoherent transition radiation. 
We do not expect any contribution from coherent radiation, because our bunch length is quite longer than optical wavelength and the use of a moderate compression with velocity bunching should avoid the formation of microbunching inside the bunch. 

Another single shot emittance measurement has been recently published \cite{barber2017measured} relying on a method developed some years ago \cite{weingartner2012}. Basically a beam produced by LWFA is focused in a triplet of permanent magnets before entering in a dispersive dipole. Due to the large energy spread (usually several $\%$), different energy parts inside the bunch are focused in different transverse dimensions.
This system works very well for large energy spread, but its main limit will appear in case of energy spread in the order of 0.5$\%$, where the difference in focus could be really small.  

\subsection{Longitudinal diagnostics}

While for transverse single shot measurements the techniques are still in developing, for longitudinal measurement there are already several possibilities that can be implemented in our machine. 
Longitudinal diagnostics is mandatory to clearly set the correct compression phase in the velocity bunching and to recover the correlated energy spread induced in this way. 
We consider to use different methods, tailoring the instrumentations to the particular machine condition. 
The single shot longitudinal phase space measurement will be performed with an X-band RF deflector (RFD), i.e a RF cavity with a transverse deflecting mode, combined with a magnetic dipole. 
The need of an X-band cavity is mainly due to the fs scale resolution obtainable in such a way.  
While this device can reach such a resolution, particular attention must be put in its design. 
The reduced iris aperture and the possibility that the beam goes out of the center inside the device, due to the transverse field, must be considered with beam dynamics simulations. 
Only one X-band RFD is operating so far at SLAC \cite{dolgashev2012rf}. 
It is designed for an energy one order of magnitude greater, so it could be used as a reference but it must be rescaled, at least in its length. 

However, for one shot not intercepting bunch length measurement, useful for instance when the beam is sent in a plasma module to correlate input and output properties of the bunch, two other systems must be implemented. 
Diffraction radiation is emitted when a charged particle passes through a hole with transverse dimension smaller with respect to the radial extension of the electromagnetic field traveling with the charge. 
Coherent emission arises when the observed wavelength is longer with respect to the bunch length. 
For our case, where this time length can be in the range between few ps and few fs, it means to have several detectors, each one sensitive to a range of wavelengths ranging from FIR to VIS light. 
This kind of measurement can be performed in multi-shot mode using a Martin Pupplet or Michelson interferometer \cite{chiadroni2013}, or in a single shot (higly desiderable) disperding the radiation and collecting it in a linear detector. 
The complete analysis of the spectrum leads to the reconstruction of the longitudinal bunch shape. 
There are already several example of such kind of measurement, using different approaches. 
One is based on a single KRS-5 (thallium bromoiodide) prism \cite{maxwell2013coherent} and another on a series of separate spectrometer working in different wavelenght scales \cite{heigoldt2015}. 
Also, in order to set the compression phase, sometimes only a relative measurement of the coherent radiation integrated on the whole bandwidth of the detector is enough. 
We foreseen to use this compression monitor in almost two positions along the machine. 
This system can be also used to monitor the phase stability of the section used for compression and to eventually stabilize it with a feedback.

Another single shot device is based on EOS (Electro Optical Sampling). The electric field co-propagating with the bunch can rotate the polarization of a laser impinging on a non linear crystal such as GaP or ZnTe. 
By using a scheme called spatial decoding \cite{pompili2014first}, realized with an angle of incidence between the probe laser and the crystal, it is possible to retrieve the longitudinal beam profile in one shot. 
The advantage of such a scheme with respect to coherent radiation is definitely that there is no reconstruction of the bunch shape starting with frequency analysis, with the problems related to the correct transport and propagation of all the wavelengths in the spectrum. 
But the disadvantage is the temporal resolution, limited by the crystal bandwidth or by the length of the laser probe. 
Typical values are in the order of 40-50 fs. 
However, this diagnostics will be very important in our machine because while the X band RFD offers a high resolution for the measurement of very short bunches, i.e. in the fs region, it will be not the best choice for ps bunch length. 
On the other end the EOS can cover easily this range of dimensions, being also not intercepting. 
Also it is often used as bunch of arrival monitor, that it is very important in some plasma acceleration schemes, as for instance external injection \cite{clayton1996}.

\subsection{Charge and trajectory diagnostics}

The control of the charge and the trajectory at a few pC and few $\mu$m is mandatory in this machine, especially in the plasma interaction region. 
About the charge, Bergoz Turbo-ICT (integrated current transformer) can be the best choice, allowing the measurement of a charge as low as 50 fC. 

Regarding the optics we are very sensitive to the beam trajectory at the entrance of every RF module, in particular the part of the machine in X band, and inside the plasma accelerator. 
Conventional stripline BPM (Beam position monitor), similar to those already in use at SPARC$\_$LAB can be considered for such a task. 
They can offer good signal to noise ratio down to few pC charge and a resolution in the order of tens of $\mu$m. 

	\begin{figure}[htb]
  \centering
  \includegraphics*[width=80mm]{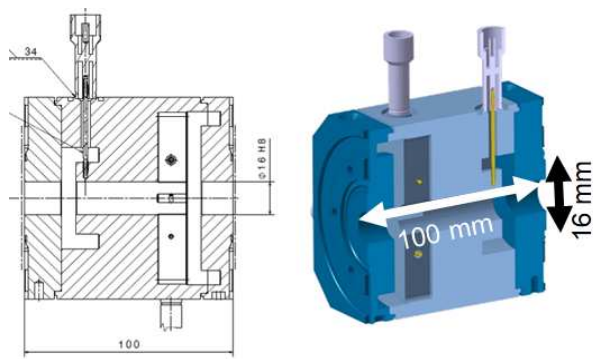}
  \caption{Compact cavity BPM developed at PSI is a great example of a compact and high resolution device.}
  \label{fig7}
\end{figure}

However, this kind of devices can be used only at the beginning of the accelerator, where the beam pipe is 40 mm. 
But starting from X band structures, the pipe size decreases.  
Also, one of the most important parameter is the length of the device. 
Due to the large number of such a system its length must be as short as possible. 
There are several studies of very compact C or X band cavity BPM \cite{keil2013design} with a total length of about 10 cm, that can be useful consider for such a task, see for instance Fig.\ref{fig5}. 
The possibility to have $\mu$m level resolution, even at very low charge (few pC), using so short longitudinal space, makes compact cavity BPM very appealing for our project, and we plan to use widely in our machine. 

\section{Conclusions}

The realization of a new compact accelerator in Frascati, based on RF X-band technology and plasma acceleration and able to drive a FEL in the soft X rays, sets a series of challenges in all of the equipment, and in particular in the diagnostics. 
Compact, single shot, high resolution devices must be implemented to properly monitor the machine. 
In preparing a conceptual design report we found that while there are already several diagnostics for trajectory, charge and longitudinal parameters, that can be adapted to our case, there is a lack of techniques for single shot emittance. 
To overcome this problem we propose a widely use of the betatron radiation to monitor the beam properties inside the plasma channel and to test some new ideas outside the plasma, but only after a capture optics, to measure the properties of the beam that will be used for applications.

\section{ACKNOWLEDGEMENTS}

This work was supported by the European Union‘s Horizon 2020 research and innovation programme under grant agreement No. 653782.


\bibliographystyle{elsarticle-num}

\end{document}